\documentclass[preprint]{aastex}
\usepackage{amsbsy} 

\newcommand{\Msun}{M_\odot}

\slugcomment{to be published in MNRAS}

\shorttitle{Externally Fed Star Formation: A Numerical Exploration}
\shortauthors{Mohammadpour \& Stahler}

\begin{document}

\title{Externally Fed Star Formation: A Numerical Study}

\author{Motahareh Mohammadpour\altaffilmark{1} and 
Steven W. Stahler\altaffilmark{2}}

\altaffiltext{1}{Department of Physics, University of Mazandaran,
Babolsar, Iran}
\altaffiltext{2}{Astronomy Department. University of California,
Berkeley, CA 94720}

\email{Sstahler@astro.berkeley.edu}

\begin{abstract}
We investigate, through a series of numerical calculations, the evolution
of dense cores that are accreting external gas up to and beyond the point of 
star formation. Our model clouds are spherical, unmagnetized configurations 
with fixed outer boundaries, across which gas enters subsonically. When we
start with any near-equilibrium state, we find that the cloud's internal 
velocity also remains subsonic for an extended period, in agreement with 
observations. However, the velocity becomes supersonic shortly before the star 
forms. Consequently, the accretion rate building up the protostar is much 
greater than the benchmark value $c_s^3/G$, where $c_s$ is the sound speed in 
the dense core. This accretion spike would generate a higher luminosity than 
those seen in even the most embedded young stars. Moreover, we find that the 
region of supersonic infall surrounding the protostar races out to engulf much 
of the cloud, again in violation of the observations, which show infall to be 
spatially confined. Similar problematic results have been obtained by all
other hydrodynamic simulations to date, regardless of the specific infall
geometry or boundary conditions adopted. Low-mass star formation is evidently
a quasi-static process, in which cloud gas moves inward subsonically until the
birth of the star itself. We speculate that magnetic tension in the cloud's
deep interior helps restrain the infall prior to this event.
\end{abstract}


\section{Introduction}

We have known for over three decades that low-mass stars form inside the dense 
cores found throughout larger molecular clouds. Yet the manner in which the 
cores themselves first arise and then evolve to the point of collapse has
eluded understanding. This situation is now changing. A growing body of 
observations, conducted by many researchers, is supplying detailed results that
will eventually be the foundation of a much more complete theoretical model.

One key result has been the detection of internal contraction within dense
cores. These objects are generally quiescent, with observed molecular line 
widths that are only marginally broader than thermal \citep{g98}, at least
until the dense core boundary is reached \citep{p10}. Spectroscopic
observations, however, reveal asymmetric emission profiles that signify 
inward, subsonic motion, even before a star is formed \citep{lmt01}. Among
starless cores, which constitute at least half the objects observed 
\citep[e.g.][]{f09}, this so-called ``infall signature'' is especially common 
in those with the highest central column density \citep{ge00}. Presumably, the 
latter are more evolved, a supposition bolstered by the observation of 
systematic trends in chemical abundances \citep{kc08}.

Even more advanced dense cores, those containing low-mass young stars, also
exhibit asymmetric line profiles. The inferred speeds in this case are still
subsonic in the outer regions of the cloud, but supersonic within the interior,
as one would expect for an object that is undergoing gravitational collapse.
Curiously, the spatial extent of supersonic infall is extremely limited, at
least in the handful of objects thus far observed with sufficient precision.
The spectroscopic studies of \citet{c95}, \citet{ge97}, \citet{df01}, and 
\citet{b02} find that the sonic transition occurs only 0.01 - 0.02~pc from the
protostar, in a dense core of linear size 0.1~pc. Submillimeter observations of
dust continuum emission reach a similar conclusion via a different route.
\citet{ser02} show that collapse cannot pervade the interior of the core, as it
would alter the density profile in a manner that is not observed.

This finding of spatially confined infall is sharply at odds with conventional 
theory. In the still widely used self-similar model of \citet{s77}, the infall 
region spreads outward as a rarefaction wave traveling at the ambient sound 
speed $c_s$. Just inside the wavefront, at \hbox{$r\,=\,r_f\,=\,c_s\,t$}, the
infall speed is zero, and becomes supersonic at 
\hbox{$r\,\simeq\,0.4\,r_f$}.\footnote{\citet{c95} and \citet{ge97} quote 
values of $r_f$ obtained by matching their data to the model of \citet{s77}. We
further used this model to infer the radius of the sonic transition.} By the 
time the protostar's mass becomes half that of the parent cloud, the 
rarefaction wave has already reached the cloud boundary. The large number of 
simulations that followed Shu's semi-analytic treatment paint a similar 
picture \citep{fc93,o99,my03,a05}.

Another departure between theory and observation is embodied in the venerable
``luminosity problem.'' The luminosity of a protostar is supplied by mass
accretion onto its surface, and is in fact proportional to ${\dot m}_\ast$,
the instantaneous mass accretion rate. \citet{k90} first noted that the
${\dot m}_\ast$ given by theory yields a luminosity greater by an order of
magnitude than those observed in the most deeply embedded infrared and
submillimeter sources. In the model of \citet{s77}, 
\hbox{${\dot m}_\ast \,=\, m_\circ\,c_s^3/G$}, where $m_\circ$ is a numerical 
factor close to unity. In numerical simulations, ${\dot m}_\ast$ varies in
time, but is generally even larger, exacerbating the discrepancy 
\citep[e.g.][Fig.~7]{o99}.\footnote{Some recent simulations find that infalling
gas landing on the protostellar disk is stored for many orbital periods, and 
released onto the star in episodic bursts. Thus, the luminosity problem is 
ostensibly averted, at least in between bursts \citep[see][and references 
therein]{vb10}. For computational convenience, these simulations treat the 
entire dense core as a thin disk, and so cannot address the issue of cloud 
infall.} 

Solving both these problems, which are almost certainly related, will require
an improved model of cloud collapse. The infall signature observed in starless
cores can sometimes be detected far outside the densest gas \citep{lmt01}. 
That is, a subsonic inward drift is superposed on the turbulent gas surrounding
the core. In some cases, inward contraction is seen throughout an entire, 
cluster-forming cloud \citep{wbm06}. These observations support the entirely 
reasonable view that there is no physical barrier between a dense core and its 
external medium. Thus, the inward creep of gas that forms the object should 
continue unabated through the point of star formation.

In conventional simulations, however, the core boundary is characterized by
zero inward velocity. Thus, the initial mass of the dense core is never 
replenished, but steadily drains onto the protostar, at least until a 
stellar wind disrupts the cloud and ends accretion. More realistically, the
core boundary should be {\it open}. Matter drifts across this boundary
continually, replenishing in part the mass being lost to the central star. The 
actual drift speed is subsonic, as indicated by the spectroscopic studies
mentioned earlier.  

Theorists have begun to consider modified collapse models that incorporate an 
open boundary. \citet{go99,go11} simulated dense core formation out of a
converging flow (spherical or plane-parallel). A dense, hydrostatic object 
grows and then eventually collapses to form a central star. The model core in
this case indeed had an open boundary, but Gong~\&~Ostriker chose the infall 
speed to be supersonic, more characteristic of the inferred {\it random}
speeds in background gas than the systematic, inward {\it drift} speed. Their
dense core thus formed inside a shock front, and was supersonically collapsing
as a whole even before the star formed. This prediction is not in accord with
the velocity distribution inferred from spectroscopy.

\citet{ds12} explored a spherical, open-boundary model, in which gas enters
subsonically. For simplicity, they removed the dense core boundary to infinity,
i.e., to a radius much greater than the accretion value, 
\hbox{$2\,G\,M_\ast/c_s^2$}, where $M_\ast$ is the (changing) protostar mass.
They also assumed that $M_\ast$ increases through a steady-state flow, in which
the mass transport rates across all spherical shells surrounding the star are
identical and equal to ${\dot m}_\ast$. Within this context, the results were
encouraging. The actual value of the accretion rate is reduced from
$c_s^3/G$ by a factor of $2\,\beta$, where $\beta$ is the incoming Mach number.
The sonic point, marking the boundary of true infall onto the star, spreads
outward subsonically, at speed $\beta\,c_s$. Both findings are closer to what is
observed. 

In this contribution, we use numerical simulations to treat fully
time-dependent collapse with an open boundary, still within the framework of a 
spherically symmetric flow. We endow the dense core with a finite radius, whose
value is taken from observations. Recent, far-infrared imaging by the 
Herschel satellite finds that dense cores tend to be nested inside narrow
filaments within the larger parent cloud \citep{m10}. Orthogonal to the
filaments are fainter striations that are roughly parallel to the ambient
magnetic field \citep{p12}. Gas may be flowing along the striations to join 
onto the filament.

The main filaments imaged by Herschel have lengths of several parsecs and 
strikingly similar diameters of about 0.1~pc \citep{a11}. This figure matches
the long-known diameters of dense cores \citep{jma99}. It is becoming clear 
that the size of a dense core is predetermined, i.e., the object forms by gas 
drifting into, and perhaps also along, a filament of fixed diameter. In our
simulations, we have taken the outer boudary of the flow to be a sphere that is
0.1~pc in diameter.

The new simulations corroborate, but only in part, the previous results from the
steady-state model. We find again that the sonic point spreads slowly and that
${\dot m}_\ast$ falls appropriately below $c_s^3/G$. However, there is an early,
transient stage of the inflow, during which ${\dot m}_\ast$ is much 
{\it greater} than $c_s^3/G$. During this ``accretion spike,'' the sonic point 
races out at a speed higher than $c_s$. Although the duration of the 
accretion spike is relatively brief, ${\dot m}_\ast$ is so high that the 
protostar gains essentially all of its mass under these conditions. Moreover, 
supersonic infall engulfs most of the dense core during the period of the 
spike. Thus, this model too is at odds with the observational facts. The idea 
of externally fed star formation remains well motivated, but the actual flow 
creating the protostar must be subsonic. It seems impossible to accomplish this
feat in a purely hydrodynamic scenario. That is, magnetic support must play a 
key role during dense core contraction.

In Section~2 below, we present our method of solution, including the manner in
which we choose the initial state of the dense core. Section~3 gives numerical
results for the collapse, focusing on the critical evolution of the mass
accretion rate and of the sonic point. Finally, Section~4 places our findings
in context, and suggests fruitful directions for future investigations.

\section{Solution Strategy}
\subsection{Numerical Method}

Following many previous researchers, we idealize the flow building up the dense
core to be spherically symmetric. Real dense cores are not spheres, but have
projected aspect ratios of about 2:1 \citep[e.g.][]{r96}. It is an open 
question whether the observed core shapes result from {\it internal} magnetic 
support, which is inherently anisotropic, or from {\it external} influences, 
such as pressure gradients across the bounding filaments. By restricting 
ourselves to a spherical inflow, we are implicitly discounting all forces but 
those associated with thermal pressure and self-gravity. This is a reasonable 
first approximation, given the narrow observed linewidths of tracer molecules. 
To the extent that our model fails to match observations, additional forces 
must be considered.  

We take the gas to be isothermal, both spatially and temporally, with a sound 
speed of \hbox{$c_s\,=\,0.2\,\,{\rm km}\,\,{\rm s}^{-1}$}. Within a molecular 
cloud, this speed corresponds to a gas kinetic temperature of 12~K, which is 
representative of observed values for starless cores \citep{jma99}. Thus far, 
our assumptions are standard ones. Our main innovation is to apply an open 
outer boundary condition. We establish the core boundary at a radius of
\hbox{$r\,=\,r_b$}, taken to be 0.05~pc. Gas drifts across this boundary at the
fixed speed $v_b$, assumed here to be 0.2~$c_s$, in accord with spectroscopic 
observations of the infall signature \citep{lmt01}. This accreting gas has a 
fixed density throughout the calculation, which we set to establish continuity 
with the initial state, as described below. We thus maintain a constant 
external rate in each evolutionary run, 
\hbox{${\dot m}_{\rm ext} \,=\, 4\,\pi\,r_b^2\,\rho_b\,v_b$}.

In all our simulations, we have employed the publicly available code
ZEUS-MP \citep{h06}. We covered the flow with 350 zones, logarithmically spaced
to give finer resolution of the central region. At \hbox{$r\,=\,r_b$}, we 
applied the inflow boundary condition, which allowed us to fix the outer 
velocity and density. At \hbox{$r\,=\,0$}, we applied the reflecting boundary 
condition, forcing the velocity to vanish at the origin. 

Once the central density climbed to 
\hbox{$6\times 10^{-14}\,\,{\rm gm}\,\,{\rm cm}^{-3}$}, some five orders of
magnitude higher than its initial value, we inserted a central sink cell, 
following the technique of \citet{bb82}. This event signaled the formation of 
the protostar. We initially chose the radius for the sink cell so that the 
density dropped by a factor of 4 from the center to its edge, and then 
maintained that radius thereafter. In practice, the sink cell covered about 8 
zones. At the edge of the cell, we applied the outflow boundary condition. We 
also calculated the stellar acccretion rate ${\dot m}_\ast$ at that point. 

\subsection{Choice of Initial State}

To see how a spherical dense core evolves while undergoing steady, subsonic
accretion, we have explored a broad range of initial states. In all cases, we
began with a linear velocity profile. That is, the initial velocity $v_0 (r)$,
taken to be positive for inflow, was chosen to be 
\begin{equation}
v_0 \,=\, v_b\,r/r_b \,\,.
\end{equation}
By adopting this profile, we ensured that the subsequent accrual of mass 
through the boundary occurred smoothly, with no artificial pileup or deficit 
in gas at $r\,=\,r_b$.

Each evolutionary run is then characterized by the starting density profile,
$\rho_0 (r)$. Table~1 lists all our runs, with the leftmost column giving the
initial density contrast from center to edge, $\rho_c/\rho_b$. In the first
two runs, we adopted a uniform initial density. Thus, $\rho_c/\rho_b$ is unity
in these cases. This imposed condition is, of course, highly artificial
astrophysically, and was chosen to compare our results with earlier collapse
studies. Thus, we retained a closed boundary condition (\hbox{$v_b\,=\,0$})
in the first run, and switched over to our standard open boundary, with
\hbox{$v_b\,=\,0.2\,c_s$}, for the second.

In all our subsequent simulations, listed below the horizontal line in the 
table, we used an open boundary condition and an initial cloud that was in 
force balance between self-gravity and thermal pressure. That is, $\rho_0 (r)$ 
corresponded to the profile of an equilibrium, isothermal sphere. As is well 
known, these equilibria comprise a one-parameter family. For each run, we fixed
the density contrast $\rho_c/\rho_b$, and then used a numerical solution of the
isothermal Lane-Emden equation to find the density profile. We obtained the 
dimensional density $\rho (r)$ by requiring that the outer boundary be located 
at \hbox{$r\,=\,r_b$}, where $r_b$ was fixed at 0.05~pc.

An equilibrium starting state is assuredly more realistic than a uniform-density
one, which is out of force balance. Nevertheless, observations give little
clue as to the appropriate initial configuration, which is therefore rather 
arbitrary in the simulations. We tested a broad range of states, corresponding 
to both stable and unstable equilibria. That is, some initial states had 
$\rho_c/\rho_b$ below the critical, Bonnor-Ebert value of 14.0, while others 
had higher values. For our lowest density contrast, we set 
\hbox{$\rho_c/\rho_b\,=\,1.45$}. Here, the cloud mass was $0.31\,\,\Msun$, and 
the number density in hydrogen molecules was $1\times 10^4\,\,{\rm cm}^{-3}$. 
This cloud would be marginally detectable using a tracer molecule such as 
NH$_3$.

As we evolved each cloud, gas flowed across the boundary at the speed of 
\hbox{$v_s\,=\,0.2\,c_s$} and a fixed density of $\rho_b$. For continuity, we
chose the latter to be the outermost density in the initial state. Thus, the
externally imposed accretion rate ${\dot m}_{\rm ext}$ varied from one run to
another, though not by a large amount. The fifth column in the table gives the 
actual values of ${\dot m}_{\rm ext}$ in units of $c_s^3/G$. Notice that 
${\dot m}_{\rm ext}$ is less than $c_s^3/G$ in every case, in accordance with 
the earlier analysis of \citet{ds12}. 

\section{Results}

\subsection{Uniform-Density Initial States}

\subsubsection{Closed Boundary}

By choosing a uniform-density initial state and setting the boundary velocity
$v_b$ to zero at all times, we are essentially repeating the classic study of
\citet{l69}. Comparison of our results with Larson's served both to validate 
our numerical method and to identify the changes effected by subsequently 
employing an open boundary. We chose the initial density $\rho_0$ to be the
minimum required for the cloud to undergo collapse. We found this minimum 
density to be 
\hbox{$\rho_0 \,=\, 1.40\times 10^{-19}\,\,{\rm gm}\,\,{\rm cm}^{-3}$},
corresponding to a cloud mass of $1.08\,\,\Msun$. Using the conventional 
definition of free-fall time,
\begin{equation}
t_{\rm ff} \,\equiv\, \sqrt{{3\,\pi}\over{32\,G\,\rho_0}} \,\,,
\end{equation}
our initial state had \hbox{$t_{\rm ff} \,=\, 1.78\times 10^6\,\,{\rm yr}$}. 
\citet{l69} adopted the slightly larger $r_b$-value of 0.053~pc, and used a 
correspondingly lower $\rho_0$ of 
\hbox{$1.10\times 10^{-19}\,\,{\rm gm}\,\,{\rm cm}^{-3}$}.  

By design, the central density of the cloud monotonically increased with time.
However, the internal velocity did not develop smoothly from the initial linear
profile. As long as the velocity was everywhere subsonic, the profile changed 
in an erratic manner. Gas did not simply come to rest at the origin, but
underwent a bounce, temporarily creating a central region of outwardly moving 
material. This reversal, in turn led to irregularities in the evolving density 
profile. At the time $t_s$ listed in the table, the peak inward velocity was 
supersonic. After that point, both the velocity and density profiles became
smooth.\footnote{The transient irregularities described here did not occur if 
we used an initial density substantially higher than the minimum value 
required for collapse.}

The left panel of Figure~1 shows the evolution of the density profile. The 
solid curves show the rise in density prior to $t_\ast$, the time at which the
protostar forms, according to the criterion previously described. In the outer
region of the cloud, the density quickly develops an $r^{-2}$ profile, and
flattens toward the center. These results qualitatively match those described 
by \citet{l69} and displayed in his Figure~1. In Larson's simulation, the star 
formed $4\times 10^5\,\,{\rm yr}$ after the initiation of collapse, while our 
corresponding time was closer to $3\times 10^5\,\,{\rm yr}$ (see Table~1). 
This difference stems from our choice of a slightly higher initial density.  
 
Following the formation of the protostar, the surrounding gas eventually enters
into a state of free-fall collapse. The mass transfer rate across each 
spherical shell is about the same, and $\rho (r)$ attains a flatter, 
$r^{-3/2}$ profile.\footnote{It is this flattening of the density profile that 
is {\it not} detected throughout the bulk of observed dense cores containing 
stars \citep{ser02}.} As cloud mass drains onto the central star, the free-fall
speed at any radius increases and the overall magnitude of the density 
declines. These later density profiles are displayed as dashed curves in the 
left panel of Figure~1. Once again, the same results are well described by 
\citet{l69} and subsequent authors. 

The fact that the peak interior velocity becomes supersonic prior to the
formation of the star has an important consequence. In Figure~2, we show the
evolution of ${\dot m}_\ast$, the mass accretion rate onto the star. This rate
is, of course, only defined for \hbox{$t\,\ge\,t_\ast$}, and has been set to
zero before that time. Shortly after this time, ${\dot m}_\ast$ is very large 
relative to the canonical value $c_s^3/G$, exceeding it by a factor of 20. The 
rate then drops in a relatively brief time. The star thus undergoes an initial
``accretion spike,'' caused by the buildup of supersonic speeds in the dense 
core prior to the formation time $t_\ast$.

To understand further the physical basis of this phenomenon, we have run 
comparison simulations in which ${\dot m}_\ast$ was calculated from 
\hbox{$t\,=\,0$}. For \hbox{$t\,<\,t_\ast$}, we used the mass transport rate 
across what later became the outer boundary of the sink cell. The resulting 
curves were indistinguishable from those in Figure~2. Thus, the sharp rise in 
the accretion spike is independent of the precise definition of $t_\ast$. 

The amount of mass accumulated during the accretion spike is not small, as we
now quantify. Observational comparison of the mass distributions of dense cores
with those of newly formed stars (the initial mass function) shows that, on 
average, a dense core produces a star containing about 0.3 times the core mass
\citep{a07,r09}. Accordingly, we let $t_f$ be the time in our simulations when 
the protostar is likely to build up its final mass, which we set to 0.3 times 
the current cloud mass. This time is given in the fourth column of the table. 
For the specific case at hand, we see that $t_f$ exceeds $t_\ast$ by a 
relatively small amount, about \hbox{$2\times 10^4\,\,{\rm yr}$}. 

The time $t_f$ is also shown by a filled circle in Figure~2. We see vividly 
that this time occurs before ${\dot m}_\ast$ has declined to a value of
$c_s^3/G$ or less. In other words, {\it most of the stellar mass accumulates
while the accretion rate is still large.} This result, as we shall see, is 
quite general. In the present case, the average accretion rate from time 
$t_\ast$ to $t_f$ is about $8\,c_s^3/G$. This average rate, symbolized as 
$\langle{\dot m}_\ast\rangle$, is given for all simulations as the sixth 
column in the table. To understand fully the physical character of the 
collapse, we have allowed this and subsequent simulations to run well past 
\hbox{$t\,=\,t_f$}, although the star would in fact have dispersed the
dense core soon after this point.

Another quantity that we can compare with observations is the sonic point
$r_s$, defined as the outermost radius at which the incoming fluid speed 
matches $c_s$. As the collapse proceeds and the gas velocity generally 
increases, $r_s$ moves outward. Figure~3 shows the advance of $r_s$ in the
first simulation. At \hbox{$t\,=\,t_s$}, the sonic point is located at
$0.14\,r_b$. It first races outward at supersonic speed. We denote this
initial speed as $\left({\dot r}_s\right)_0$ and list it in the last column of
the table. Although its outward advance slows with time, the sonic point is 
still traveling at $1.03\,c_s$ at the time $t_f$, again shown by the filled 
circle in the figure. At this point, $r_s$ has covered a radius of about 
$0.3\,r_b$. By the last time shown in Figure~3, the speed of the sonic point 
has declined to $0.08\,c_s$. The mass of the cloud has fallen to 0.08 times 
its initial value. 

\subsubsection{Open Boundary}

Our next initial state was again of uniform density, and with the same
$\rho_0$-value. This time, however, we applied the more realistic open
boundary condition at the cloud edge. The righthand panel of Figure~1 shows that
the density profiles evolve in a qualitatively similar manner as for the
closed boundary. Prior to the formation of the central star, which now occurs
at the earlier time of \hbox{$t_\ast\,=\,1.86\times 10^5\,\,{\rm yr}$}, each
profile again consists of an $r^{-2}$ envelope and a relatively shallow central
core. After the star forms, the density attains the $r^{-3/2}$ slope 
characteristic of steady, free-fall collapse. In this case, the cloud mass
never vanishes, and it takes longer for the stellar mass to overtake it. Hence,
it also takes longer for the infall speed to approach true free-fall, and the
density profiles accordingly evolve more slowly.

The profiles in Figure~1 have one curious feature that calls for explanation.
Although gas flows across the boundary with a fixed density, the figure shows
that $\rho (r)$ just inward of that point falls sharply before starting its
more gradual, inward rise. This sudden drop in density, typically by a factor of
2, is accompanied by a corresponding increase of the velocity. Gravity is 
pulling gas inward faster than it can be supplied from the outside, creating a
partial vacuum. This phenomenon is unphysical, an artifact of our fixing the 
drift speed at $0.2\,c_s$. Since the exterior mass accretion rate is not
effected, there is no impact on the collapse process. In any event, we shall 
later see that the same phenomenon occurs, albeit to a lesser extent, when 
we begin with states near hydrostatic equilibrium.

Returning to Figure~2, which shows the evolution of ${\dot m}_\ast$, we see
first that formation of the star occurs earlier when we use an open boundary.
The reason is simply that more cloud mass is available to accumulate at the
origin. As before, the internal velocities become supersonic, and 
${\dot m}_\ast (t)$ exhibits a sharp initial spike. From the
location of the filled circle, marking the time $t_f$, we again see that the
bulk of the protostar's mass is gained while the accretion rate is still
relatively high. According to Table~1, the average rate building up the 
protostar is $10.9~c_s^3/G$. In this case, ${\dot m}_\ast (t)$ does not 
asymptotically vanish, as it did for the closed boundary, but levels off at the
externally imposed value, \hbox{${\dot m}_{\rm ext} \,=\, 0.84\,c_s^3/G$}, as 
also listed in the table.

Finally, Figure~3 shows that the sonic point starts at a much larger radius
($0.54~r_b$) than in the closed boundary case ($0.14~r_b$). That is, more of the
cloud is collapsing supersonically at the point of star formation. The initial
speed of the sonic point, $\left({\dot r}_s\right)_0$, is again quite high,
almost $2\,c_s$. At \hbox{$t\,=\,t_f$}, $r_s$ covers 0.64 of the full cloud
radius. The point comes close to the cloud boundary by
\hbox{$t\,=\,3\times 10^6\,\,{\rm yr}$}, and thererafter remains nearly static.
That is, almost the entire cloud is collapsing supersonically at this late
epoch, despite the continuing, subsonic injection speed.

\subsection{Equilibrium Initial States} 
\subsubsection{Density and Velocity Evolution}

We next began with configurations that are in a state close to hydrostatic
balance. As before, the density profiles did correspond precisely to 
equilibria. Since each cloud had a non-zero velocity profile from the star, 
given by equation~(1), it was actually in a near-equilibrium state that evolved 
quasi-statically, at least early in the simulation.
 
When we started with our minimal, center-to-edge density contrast of 
\hbox{$\rho_c/\rho_b\,=\,1.45$}, we found that the cloud not only evolved 
through all the stable equilibria in the isothermal sequence
(\hbox{$\rho_c/\rho_b\,<\,14.0$}), but beyond this point into the regime of
unstable states. When the central density climbed to $4\times 10^3$ times the
edge value, the peak inward velocity became supersonic. This event occurred
at \hbox{$t_s\,=\,1.54\times 10^6\,\,{\rm yr}$}. A short time later, at
\hbox{$t_\ast\,=\,1.55\times 10^6\,\,{\rm yr}$}, the central star formed, and 
both the density and velocity increasingly resembled those corresponding to
free-fall collapse onto the central mass.

The sequence of events was qualitatively the same when we started with
configurations having a larger density contrast. Figure~4 shows the evolution 
of the density for the last stable starting state, that with 
\hbox{$\rho_c/\rho_b \,=\, 14.0$}. In the purely hydrodynamic, spherical
model assumed here, this configuration is the most realistic starting state,
as recognized by many others \citep{h77,fc93,o99,a05}. However, previous 
authors increased the initial density everywhere above the equilibrium value
to ensure that the system evolved with time. We do not require this artifice, 
as the continual addition of mass from the outside naturally drives the cloud 
into collapse.

Figure~4 shows vividly how the central density shoots up before the internal
velocity becomes supersonic at \hbox{$t_s\,=\,2.52\times 10^5\,\,{\rm yr}$}.
The dashed curve is the profile at \hbox{$t\,=\,3.0\times 10^5\,\,{\rm yr}$},
after the point of star formation, 
\hbox{$t_\ast\,=\,2.59\times 10^5\,\,{\rm yr}$}. Already at this time, the 
density is everywhere diminishing, as the cloud enters a state of free fall 
onto the growing protostar. 

The evolution of the internal velocity profile, for the same starting state,
is displayed in Figure~5. Here, the times are the same as in the preceding 
plot. As long as the velocity is wholly subsonic, its peak position moves
steadily inward. As we have seen, the peak velocity becomes supersonic shortly
before the star forms. From the shape of the velocity profile, it is clear
that there are {\it two} sonic points just after $t_s$. The inner one vanishes
once the star forms and the flow assumes the typical free-fall profile. In
all plots of $r_s$, we only display the outer point. Note also that the time 
$t_s$ when the flow becomes supersonic is 3.45 times the cloud's free-fall 
time, based on its initial central density. \citet{sy09}, who tracked the cloud 
evolution from the same starting state via perturbation theory, reached a 
similar conclusion. Both results indicate that a starless core can contract 
subsonically for a relatively long time before entering into a state of true 
collapse.

Two final remarks are in order. First, Figure~5 shows that the velocity rises
quickly above its imposed boundary value just inside of $r_b$. This rise
corresponds to the drop in outer density mentioned earlier, and occurs here
for the same reason. Lastly, we see that no shock arises in the flow during
the time interval from $t_s$ to $t_\ast$, when the velocity peaks at a 
supersonic value and then plunges inward. The reason is that even the 
supersonic gas is moving rapidly enough that it cannot be overtaken by the 
more rapidly moving gas outside it. 

\subsubsection{Mass Evolution}

Figure~6 shows the evolution of the stellar accretion rate ${\dot m}_\ast$, both
for the initial equilibrium state with \hbox{$\rho_c/\rho_b\,=\,1.45$} and
for that with \hbox{$\rho_c/\rho_b\,=\,14.0$}. Not surprisingly, the star forms
more quickly when the initial density contrast in the cloud is greater. In
both cases, there is a sharp accretion spike caused by the buildup of supersonic
velocities at early times. The total mass of the star is again accumulated
while ${\dot m}_\ast$ is relatively high, as indicated by the filled circles
in the figure. According to the table, the average ${\dot m}_\ast$ during the
epoch of star formation varies between 10 and 11 times $c_s^3/G$ for all density
contrasts of the parent cloud. The accretion rate does not decline to zero at 
long times, as in the closed-boundary simulation, but approaches the externally
imposed value. All these external values fall below $c_s^3/G$, as listed in the
table.

Figure~7 shows directly the evolution of the both the cloud and stellar mass,
again for initial density contrasts of \hbox{$\rho_c/\rho_b\,=\,1.45$} and
14.0. Before the point of star formation, the cloud mass ({\it dashed curve})
climbs linearly with time, in accordance with the constant external mass
accretion rare being imposed. Just at \hbox{$t\,=\,t_\ast$}, this mass falls
sharply. The stellar mass ({\it solid curve}) climbs quickly during the 
accretion spike, then settles to the slower rate corresponding to 
${\dot m}_{\rm ext}$.

The filled circles in Figure~7 show how quickly the full stellar mass is
attained. From Table~1, the interval over which the star builds up,
\hbox{$t_f\,-\,t_\ast$}, first {\it decreases} with higher initial density
contrast, then {\it increases}. The minimum time interval occurs for an 
initial density contrast near \hbox{$\rho_c/\rho_b\,=\,8$}. Not coincidentally,
this configuration has the highest average accretion rate, 
\hbox{$<{\dot m}_\ast>$}, of all those we considered. In any event,
\hbox{$t_f\,-\,t_\ast$} is always a small fraction, typically 10~percent, of the
cloud's initial free-fall time.

The interval \hbox{$t\,-\,t_\ast$} is a similarly small fraction of the total
time elapsed since subsonic infall begin. Although we do not trace the evolution
prior to this point, our results imply that only a small fraction of dense cores
should contain protostars. And yet the observed numbers of starred and starless
cores are comparable \citep[e.g.][]{jw00}. Some of these objects might not be 
true protostars, but more mature, pre-main-sequence stars that have not yet 
fully dispersed their parent dense cores. Additionally, there is undoubtedly an 
observational bias toward observing cores of higher column density and density 
contrast \citep[see also][who emphasize this point]{go99}.

\subsubsection{Sonic Point}

We display in Figure~8 the evolution of the sonic point $r_s$ over the full
range of initial density contrasts. In all cases, this point first appears
well outside the star. Its initial speed of advance, listed as 
$\left(\dot r_s\right)_0$ in the table, is always highly supersonic. The speed
thereafter generally declines. For relatively low initial density contrast, the
point advances faster as we consider denser initial states. This trend reverses
at \hbox{$\rho_c/\rho_b \,\sim\,8$}. For even higher initial density contrast, 
$r_s$ starts out progressively closer to the star and does not advance as far 
in a given time.

In all our simulations, the trajectories \hbox{$r_s (t)$} are remarkably
similar. Figure~9 highlights this point. Here, we have overlaid all the curves
from Figure~8, shifting the vertical axis so that the sonic point always
starts in the same position. While there are detailed differences from case to
case, the same pattern recurs -- an early advance in $r_s$, followed by a 
longer expansion at slower speed. It is also generally true that $r_s$ spans 
a large fraction of the full cloud radius at \hbox{$t\,=\,t_f$}, when the star 
has accumulated most of its mass. Typically, \hbox{$r_s/r_b\,=\, 0.3\,-\,0.5$} 
at this point. For the most realistic starting state, that with 
\hbox{$\rho_c/\rho_b\,=\,14.0$}, this ratio is \hbox{$r_s/r_b\,=\,0.45$}.

The state with the largest initial density contrast, 
\hbox{$\rho_c/\rho_b\,=\,100$}, represents a highly unstable configuration,
and is not a realistic candidate model for a starless core. Indeed, the density
profile approximates that of the singular isothermal sphere, whose collapse
was studied by a number of investigators \citep{l69,p69,s77,h77,ws85}. All 
these researchers focused on {\it self-similar} collapse, a process that could 
occur, at least in principle, if the cloud's bounding surface were removed to 
infinity. On the other hand, \citet{ds12} studied the {\it steady-state} 
collapse of the singular configuration, in which the cloud is being fed by an 
external flow with subsonic speed $\beta\,c_s$, where \hbox{$\beta\,<\,1$}. In 
contrast to the self-similar studies, they found that the sonic point advances 
slowly, also at a speed of $\beta\,c_s$, while ${\dot m}_\ast$ is a constant, 
given by \hbox{$2\,\beta\,c_s^3/G$}. 

In our fully time-dependent simulation of the \hbox{$\rho_c/\rho_b\,=\,100$}
state, both predictions of \citet{ds12} are borne out, but only after initial
transients have died away. According to Table~1, the sonic point first races
out supersonically, as in all other simulations. Eventually, the speed of
this point does decline to the expected subsonic value, but only long after 
the time $t_f$. Figure~10 shows that ${\dot m}_\ast (t)$ follows a similar
route. We first see the usual accretion spike at \hbox{$t\,=\,t_\ast$}. 
Thereafter, the accretion rate does asymptotically approach the predicted value
of \hbox{$2\,\beta\,c_s^3/G\,=\,0.40\,\,c_s^3/G$}. The growth of the stellar 
mass itself, shown in Figure~10 as the dashed curve, thus slows down over this 
long epoch. Once again, the observationally relevant stellar mass is built up 
much earlier, during the early decline of the spike. The average accretion rate 
during this period thus has the relatively high value of
\hbox{$\langle{\dot m}_\ast\rangle\,=\,8.20\,\,c_s^3/G$}.

\section{Discussion}

\subsection{Onset of Collapse}

The results of this study bear not only on star formation per se, but 
also on the more general issue of starless core dynamics. As mentioned earlier,
the denser subclass of these clouds exhibits spectroscopic signs of slow
contraction. The origin of this contraction is still uncertain. Most resarchers
have concluded that ambipolar diffusion of the cloud's internal magnetic field
is too slow a process \citep[e.g.][]{lm11}. To be sure, model results 
are sensitive to the assumed cloud geometry and initial mass-loading of the 
magnetic field lines \citep{t12}. Ambipolar diffusion is also enhanced, at 
least in principle, through turbulence \citep{fa02}, but dense cores are noted 
for their quiescent interiors. All told, this mechanism does not seem 
promising. 

Since the inward motion appears only in the starless cores of highest density, 
self-gravity must be its main driver, whether or not a magnetic field is 
present. To illustrate this point, \citet{sy09,sy10} reproduced the observed 
velocity profiles using a non-magnetic, spherical cloud with the marginally 
stable density contrast of \hbox{$\rho_c/\rho_b\,=\,14.0$}. In their model,  
contraction is due to non-linear growth of the cloud's fundamental oscillation 
mode, which has zero frequency to linear order. One noteworthy finding was that 
this period of contraction is relatively protracted, typically lasting about 4 
times the cloud's free-fall time, based on its initial central density. Thus, 
the relatively high fraction of starless cores also finds a natural explanation.

Our current simulation corroborates these results. For a cloud with the
density contrast in question, the internal velocity is appropriately subsonic
for a relatively long period, and does not become supersonic until 
\hbox{$2.52\times 10^5\,\,{\rm yr}\,=\,3.45\,\,t_{\rm ff}$} after the start
of the evolution. This continued agreement with the spectroscopic studies, 
together with the proliferating observations of the immediate dense core 
environment, reinforce our belief that the contraction represents the 
approach to collapse of a cloud that has been driven to a marginally stable
state through external accretion of gas.

\subsection{Accretion Spike}

On the other hand, the conundrum known as the luminosity problem remains. We
find that, following the slow contraction phase, supersonic velocities arise
throughout the interior portion of the cloud. This fast, incoming material
creates a sudden burst of accretion at the onset of star formation. The 
accretion spike lasts less than a free-fall time, but long enough to build up
essentially all of the protostar's mass. It is only long {\it after} this
point, when a real dense core would have already been dispersed by a stellar 
wind, that ${\dot m}_\ast$ declines to the relatively slow rate found by
\citet{ds12} for a cloud undergoing steady, external accretion.

The accretion spike occurs in theoretical models regardless of the specific 
boundary conditions adopted. Thus, \citet{fc93} posited a spherical cloud of 
constant mass, whose boundary was held at fixed pressure. Their cloud radius, 
unlike ours, shrank during the evolution. Starting with the density contrast 
\hbox{$\rho_c/\rho_b\,=\,14.0$}, they found that 44\% of the cloud mass was 
moving inward supersonically at the onset of star formation, whereupon
${\dot m}_\ast$ spiked. According to their Figure~3a, the peak mass inflow
rate was \hbox{$9.3\,\,c_s^3/G$}, measured at \hbox{$r/r_b\,=\,0.047$}. The
infall rate dropped by half in a time of interval of \hbox{$0.62\,\,t_{\rm ff}$}.

For comparison, our peak accretion rate, measured much closer to the cloud
center \hbox{$(r/r_b\,=\,0.0026)$}, was \hbox{$22\,\,c_s^3/G$}, and fell by 
half \hbox{$7.5\times 10^3\,\,{\rm yr}$} after $t_\ast$, corresponding to 
\hbox{$0.10\,\,t_{\rm ff}$}. Note that both decay times are very brief compared 
to the free-fall time based on the cloud's {\it mean} initial density, which is 
\hbox{$1.7\times 10^5\,\,{\rm yr}$}. Any estimate of the average mass accretion
rate building up the protostar based on the latter definition of $t_{\rm ff}$ 
would be grossly inadequate.

The accretion spike is also not an artefact of the spherical geometry imposed
by us and other authors. \citet{go11} simulated in three dimensions the collapse
of a plane-parallel slab built up by supersonic, converging flows. Dense
condensations appeared within the postshock structure, and their evolution was
qualitatively the same as those created through spherical inflow \citep{go99}.
In particular, the authors reported a rapid rise in each core's central density
once the structure became self-gravitating, and a concurrent buildup of inward,
supersonic motion. Although they could not follow the evolution past the point
of star formation, the preceding dynamical evolution leaves no doubt that
${\dot m}_\ast$ would again have started out relatively high.

\subsection{Quasi-Static Evolution}

In summary, an accretion spike occurs whenever a self-gravitating cloud evolves
to the point of inward collapse, regardless of the detailed initial or boundary
conditions. Both the buildup of supersonic velocities throughout the bulk of the
cloud and the consequent accretion spike are sharply at odds with observations.
Together, observation and theory are forcing us to conclude that {\it low-mass 
star formation is a quasi-static process.} The parent cloud contracts slowly, 
even after it develops a large density contrast. Supersonic velocities 
naturally do appear after the star forms. The infall region surrounding the 
star subsequently spreads outward, but again in a slow, subsonic manner. 

No purely hydrodynamic model, including ours, has yielded this outcome. Some 
other force, in addition to the thermal pressure gradient, must be resisting 
self-gravity and preventing a more prompt and global collapse. Perhaps the
protostellar wind, which eventually disperses the entire cloud, helps to 
impede infall at earlier times. If so, the detailed mechanism is far from
clear \citep[see][for studies of the wind-infall interaction]{wis98,wis03}. 
The effective pressure associated with turbulence is not a candidate, as the 
nonthermal velocities inside dense cores are subsonic \citep{bg98}. We are left 
to reconsider the core's magnetic field.

According to \citet{c12}, the mass to flux ratio in dense cores, both starred 
and starless, is supercritical (see his Fig.~7). That is, the actual cloud 
mass, $M_{\rm cl}$, exceeds $M_\phi$, the magnetic critical value:
\begin{equation}
M_{\rm cl} \,>\, M_\phi \,\equiv\,{1\over{2\pi}}{\Phi\over G^{1/2}} \,\,,
\end{equation} 
where $\Phi$ is the magnetic flux threading the cloud. Here, we have followed
\citet{c12} and used the expression for $M_\phi$ from \citet{nn78}. Since 
starless cores are quasi-statically evolving prior to collapse, $M_{\rm cl}$
must still be less than the {\it total} critical mass, which includes also the
thermal component, i.e., the Jeans mass \citep[][Section~9.4]{sp04}. 

Over the relatively long, pre-collapse epoch, there is indeed time for ambipolar
diffusion to act, especially in the core's densest, central region. The most
detailed numerical simulations of this process have focused, for simplicity, on
thin-disk models of dense cores. In the still widely cited work of \citep{cm94},
the authors found that the central number density doubles in less than 1~Myr, 
once it has already reached values exceeding a few time 
\hbox{$10^4\,\,{\rm cm}^{-3}$} (see their Fig.~1). Naturally, the theoretical
models are greatly oversimplified, not only in their assumption of a thin-disk
geometry, but in their neglect of external turbulence and accretion. The gap
between theory and observation is highlighted by the observations of
\citet{c09}, who find that the mass loading of field lines in the envelopes
surrounding four dense cores exceeds that in their interiors, contrary to
conventional wisdom.

Current three-dimensional simulations of cloud collapse and star formation 
neglect any prehistory of quasi-static settling. Instead, they typically start
with a cloud threaded by a uniform and relatively weak field. Under these 
circumstances, the cloud collapses promptly and drags in flux as it forms the 
star. While this alleviates the discrepancy with observations just cited, it 
also leads to strong magnetic braking that prevents disk formation 
\citep[e.g.][]{k12}. The next frontier in theory is more careful modeling of
magnetized dense cores that undergo subsonic accretion as they contract, 
leading to the formation of a central star under more quiescent conditions 
than previously envisioned. 

\acknowledgments
We thank the referee for numerous comments that helped improve the final
manuscript. This project was carried out while M.~M. was a Visiting Scientist 
in the Berkeley Astronomy Department. She warmly thanks the Department for 
their hospitality during this productive, six-month period. S.~S. was partially 
supported by NSF~Grant~0908573.

%% the file, uncomment the \plotone, \plottwo, and
\clearpage

\begin{table}
\begin{center}
\tablewidth{8pt}
%\caption{To be supplied}
\begin{tabular}{cccccccccc}
\tableline
& \multicolumn{8}{c}
{$ \textbf{{\rm Collapse}}~~\textbf{\rm Simulations} $} &
\phantom{abc}\\
\tableline & $\rho_{\rm c}/\rho_{\rm b}$ & $t_s$ & $t_{\ast}$ & $t_f$ & 
$\dot{m}_{\rm ext}$&$<\dot{m}_{\ast}>$ & $(\dot{r}_s)_0$&\\ 
& $ $ & $(10^5~{\rm yr})$ & $(10^5~{\rm yr})$ & $(10^5~{\rm yr})$ &
$({c^3_s}/{G})$&$({c^3_s}/{G})$ & $({c_s})$&\\
\tableline 
& 1.00 (closed)& 2.86& 2.89& 3.10& 0.00& 8.35&  3.80\\
& 1.00 (open)~~~&   1.64& 1.86& 2.08& 0.84& 10.9&  1.99\\ 
\tableline \hsize=0.01pt 
& 1.45&   15.4& 15.5& 15.7& 0.35& 10.1&  4.03\\
& 2.00&   8.55& 8.62& 8.82& 0.52& 10.7&  3.54\\
& 4.00&   4.31& 4.39& 4.59& 0.66& 11.2&  3.46\\
& 8.00&   3.02& 3.11& 3.30& 0.64& 11.4&  3.48\\
& 14.0&   2.52& 2.59& 2.79& 0.59& 11.2&  3.56\\
& 20.0&   2.28& 2.35& 2.56& 0.55& 10.8&  3.70\\
& 40.0&   1.92& 1.98& 2.19& 0.48& 9.95&  3.82\\ 
& 100~ &   1.53& 1.57& 1.81& 0.40& 8.20&  4.07\\
\tableline
\end{tabular}
\caption{Parameters and selected results of the simulations. From left to 
right, the columns are:
center-to-edge density contrast of the starting state; time when the peak
contraction speed becomes supersonic; time when the protostar forms; time when
the protostar attains 0.3 times the current cloud mass; externally imposed
mass accretion rate; average mass accretion rate onto the protostar from 
$t_\ast$ to $t_f$; outward speed of the sonic point at time $t_\ast$.}
%% Any table notes must follow the \end{tabular} command.
\end{center}
\end{table}

\clearpage

\begin{figure}
\plotone{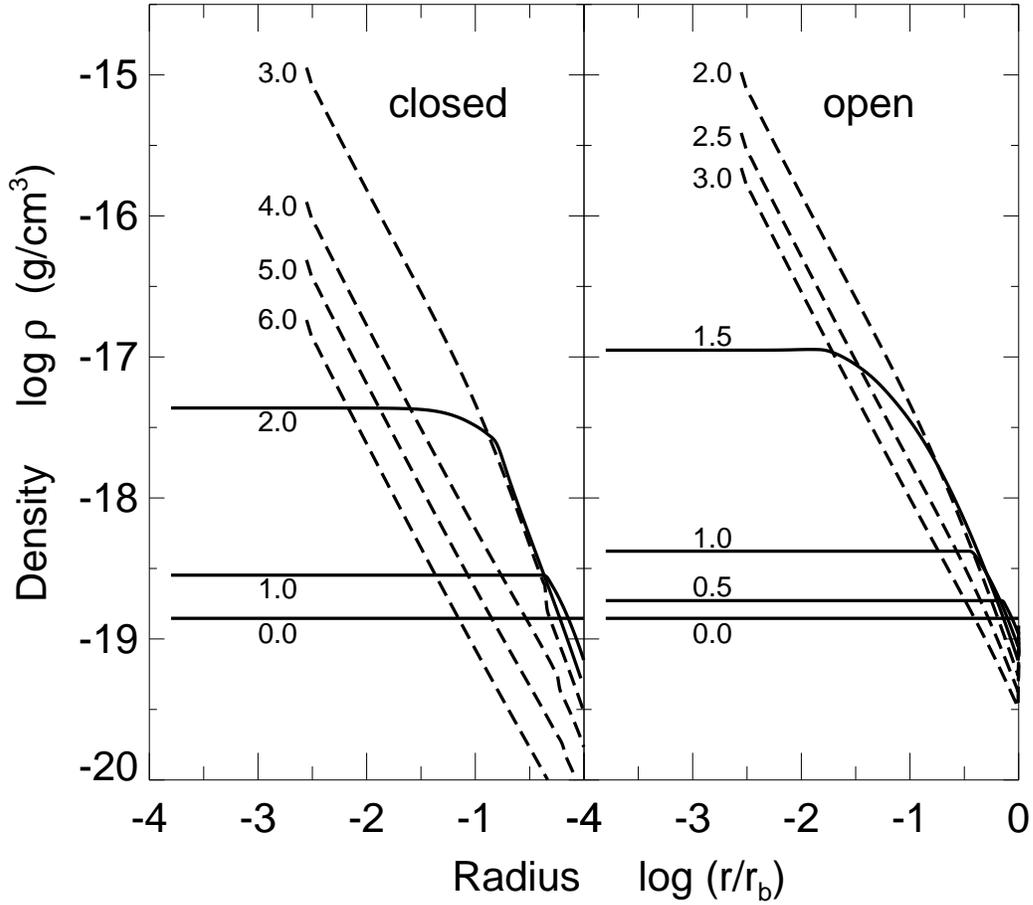}
\caption{Evolution of the cloud density profile for uniform-density starting
states. Left and right panels show the results for the closed and open boundary
conditions, respectively. In both cases, the dashed curves are the profiles 
after the protostar forms. All curves are labeled by the time, in units of 
$10^5~{\rm yr}$.}
\end{figure}

\begin{figure}
\plotone{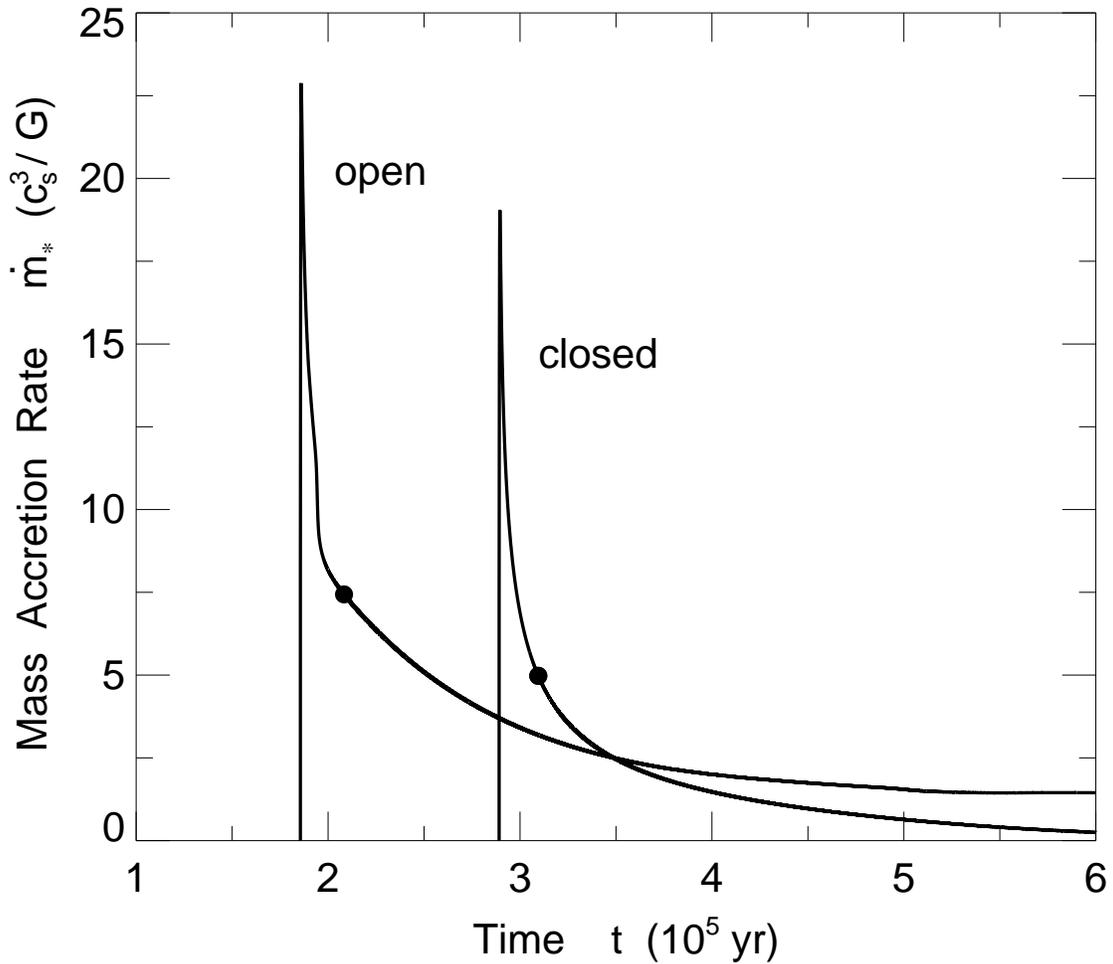}
\caption{Evolution of the mass accretion rate onto the protostar, for both
simulations using uniform-density starting states. The rate is shown in units
of $c_s^3/G$. The filled circles on both curves mark the time $t_f$, when the 
central star attains its full mass.}
\end{figure}

\begin{figure}
\plotone{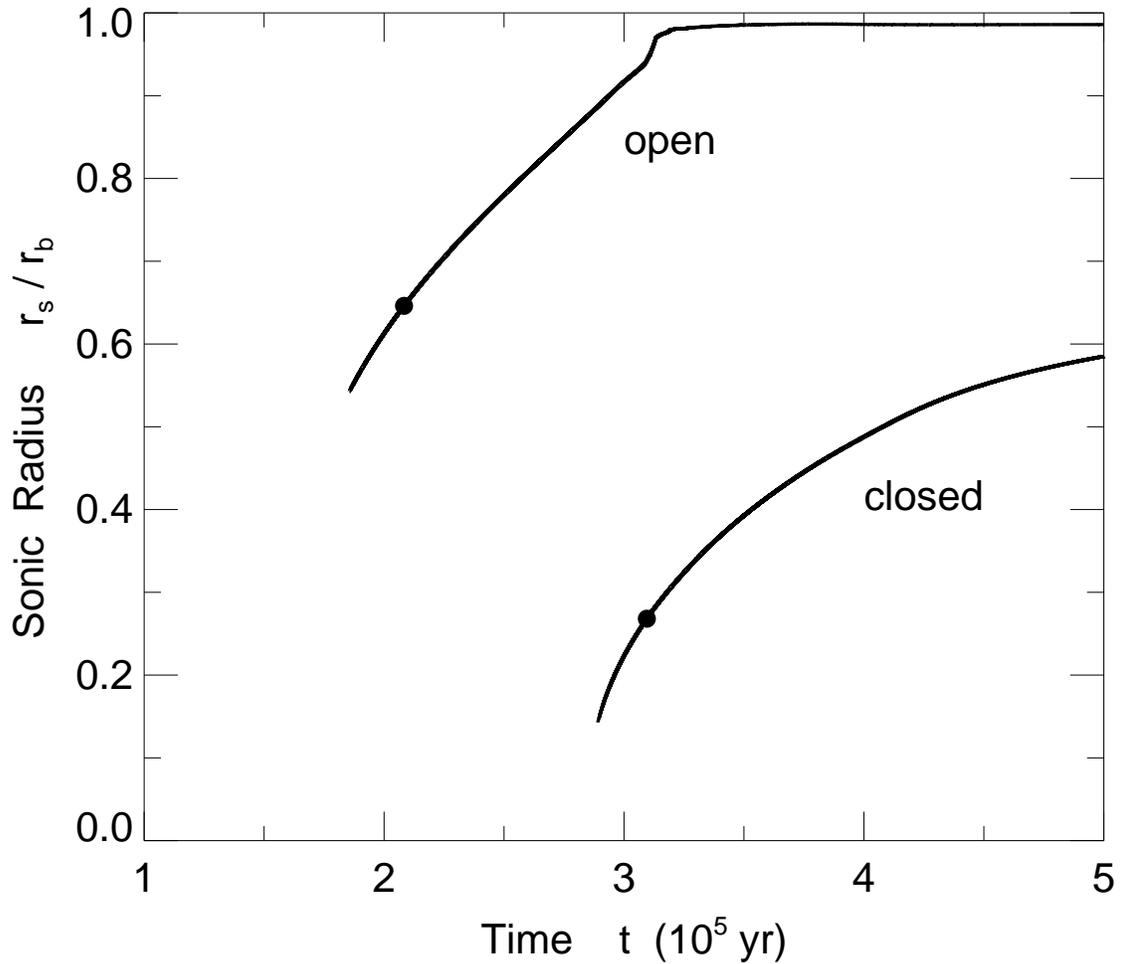}
\caption{Expansion of the sonic radius, for the simulations using 
uniform-density starting states. Plotted is the sonic radius $r_s$ relative
to the cloud boundary $r_b$. The closed circles again mark the time $t_f$.}
\end{figure}

\begin{figure}
\plotone{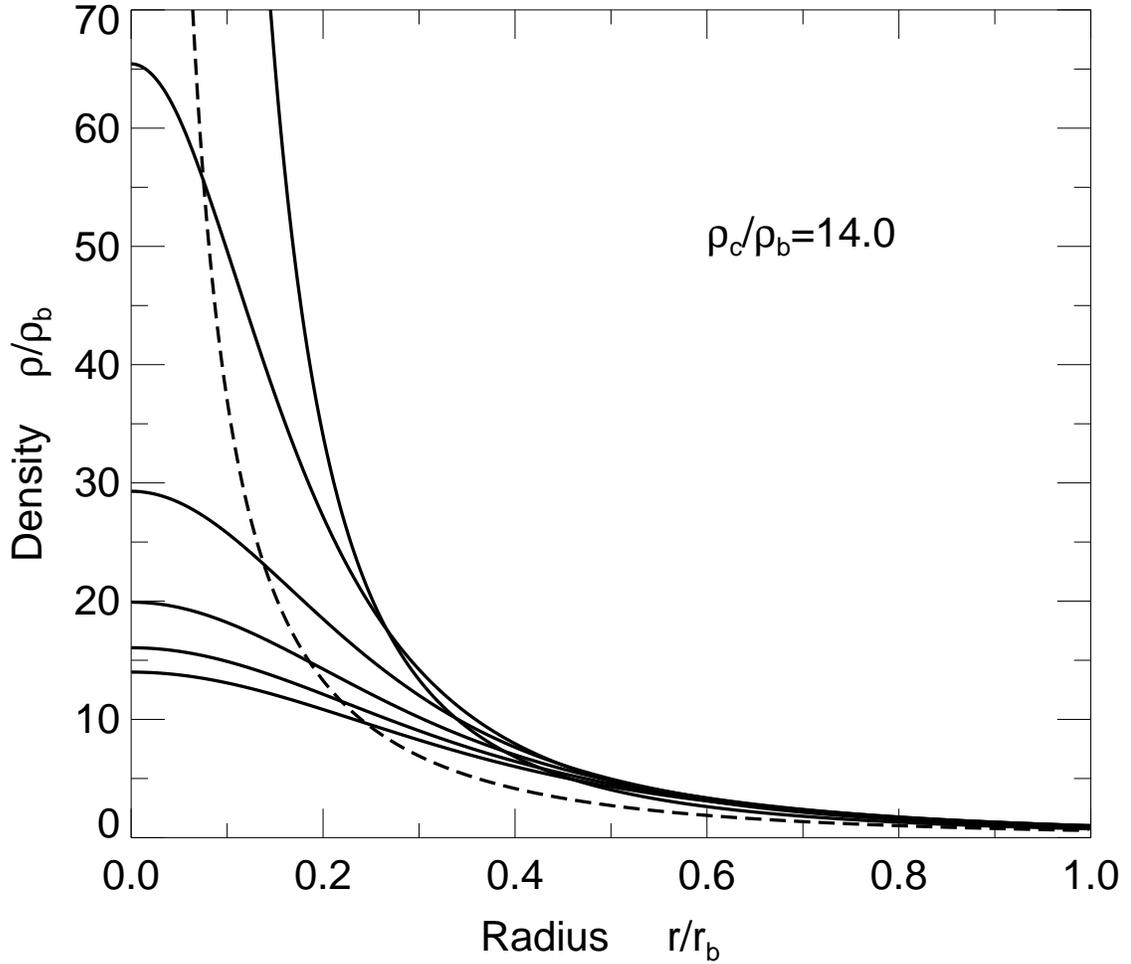}
\caption{Evolution of the cloud density profile for a starting state with a
center-to-edge density contrast of 14.0. The central density climbs 
monotonically with time. The solid curve with the lowest central density
represents \hbox{$t\,=\,0$}, and successive profiles are separated by
\hbox{$\Delta t\,=\,5\times 10^4\,\,{\rm yr}$}. The dashed curve corresponds to
the latest time, \hbox{$t\,=\,3.0\times 10^5\,\,{\rm yr}$}, which is after the 
point of star formation.}
\end{figure}

\begin{figure}
\plotone{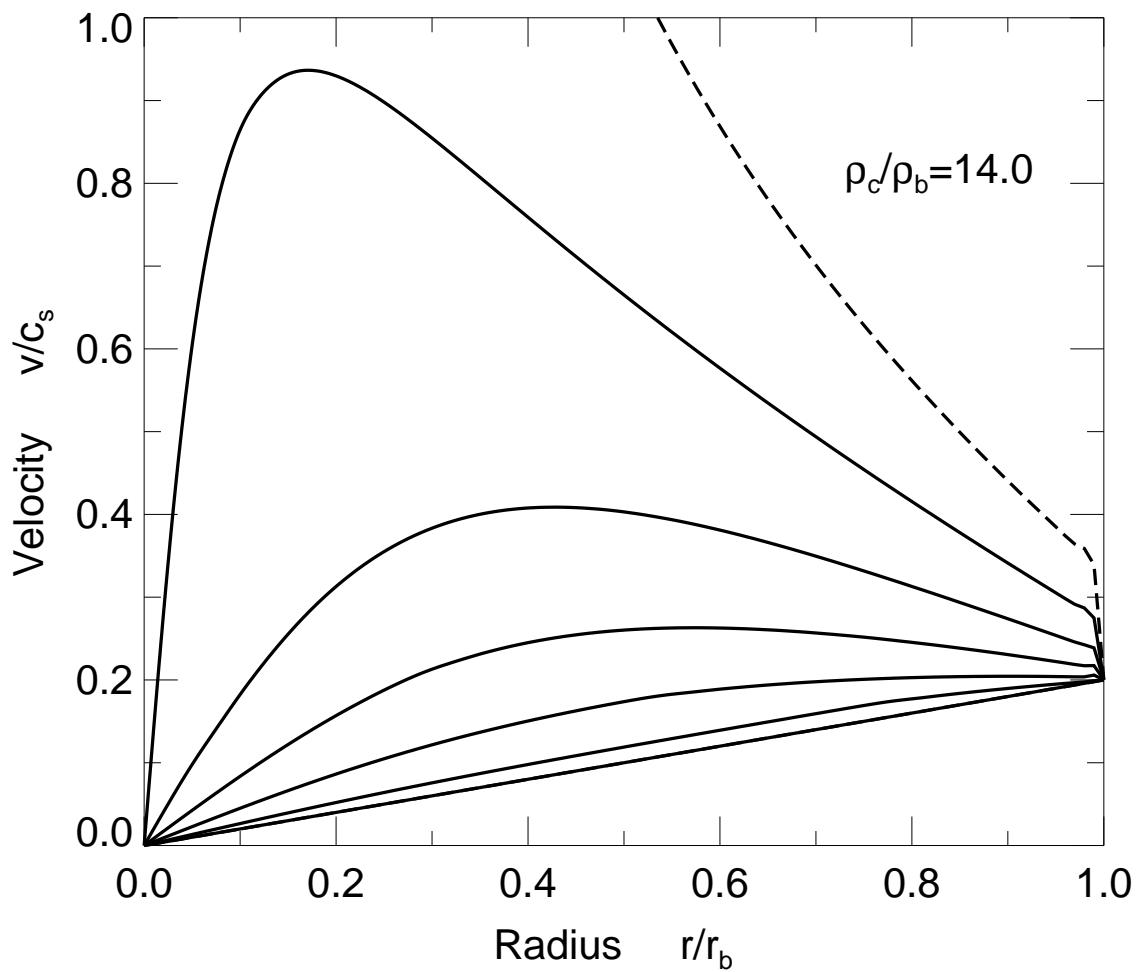}
\caption{Evolution of the velocity profile for the starting state with
\hbox{$\rho_c/\rho_b\,=\,14.0$}. From bottom to top, each curve represents the
same temporal sequence as in Figure~4. Again, the dashed curve is the only
velocity profile after the star forms. Notice, in all profiles, the jump in
velocity just inside the outer boundary.}
\end{figure}

\begin{figure}
\plotone{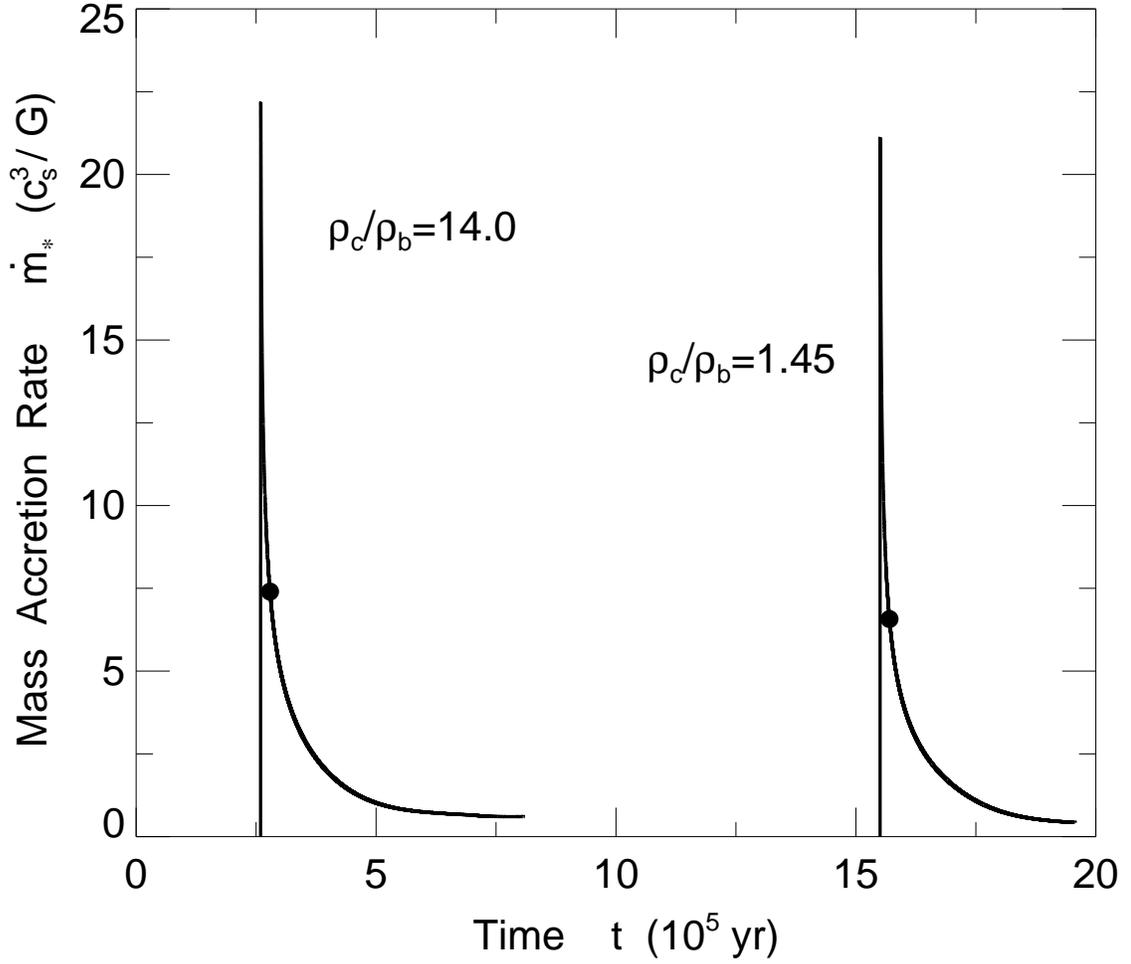}
\caption{Evolution of the stellar mass accretion rate. The two starting states
are both nearly in hydrostatic equilibrium, with the indicated density
contrasts. The filled circles again mark the time $t_f$.}
\end{figure}

\begin{figure}
\plotone{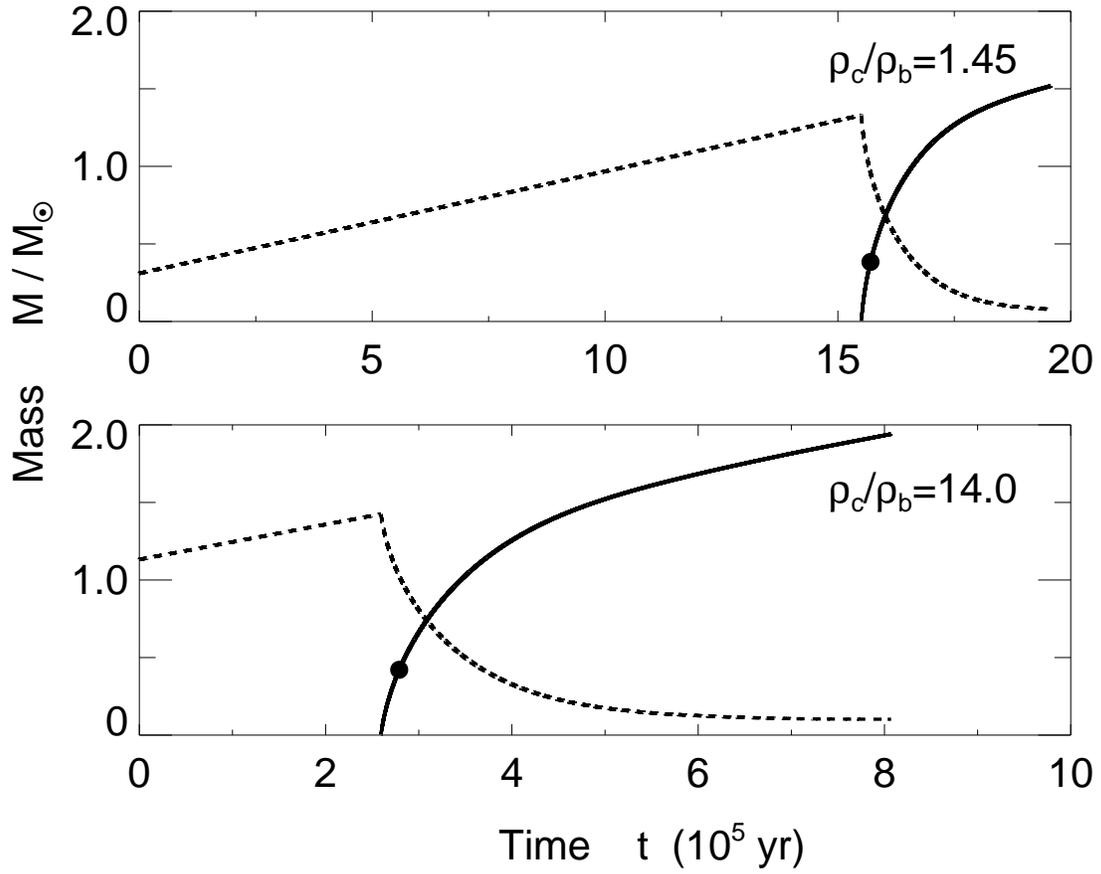}
\caption{Evolution of the stellar mass ({\it solid curve}) and the mass of the
dense core ({\it dashed curve}). The starting density contrast of the cloud is
given in each panel. The filled circles again mark the time $t_f$.}
\end{figure}

\begin{figure}
\plotone{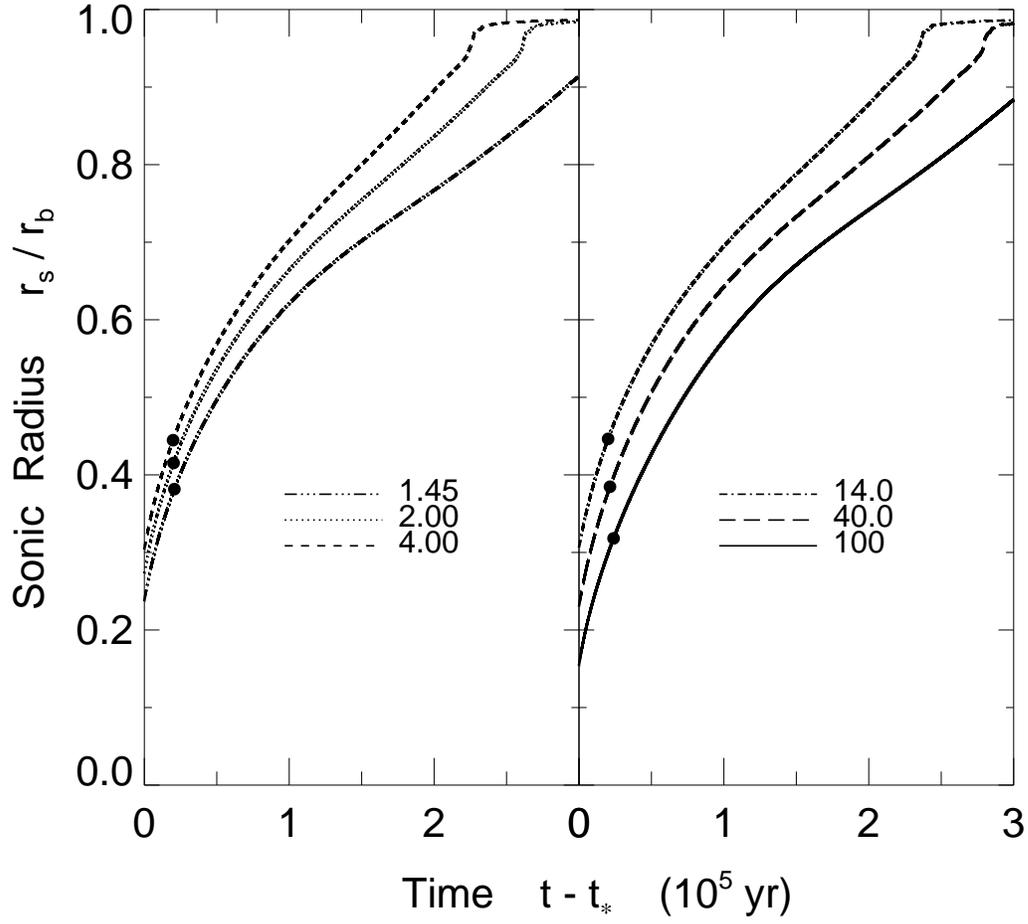}
\caption{Expansion of the sonic point. Plotted is the position of $r_s$, 
normalized to the boundary radius $r_b$, as a function of time elapsed since
star formation. Each curve corresponds to the initial density contrast 
indicated. The filled circles mark the time $t_f$}
\end{figure}

\begin{figure}
\plotone{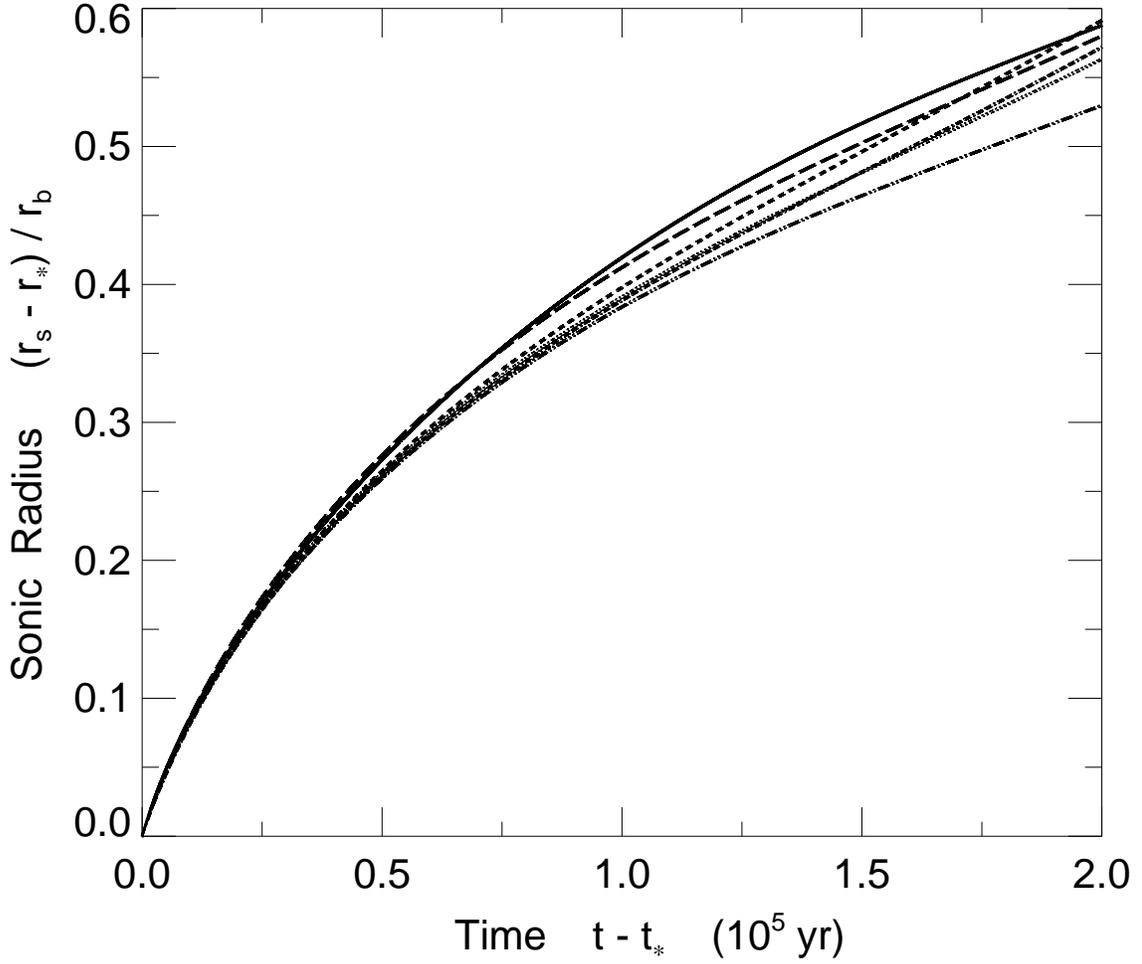}
\caption{Trajectory shapes for the sonic point. Displayed are all the curves
of $r_s (t)$ from Figure~8, but renormalized to that they begin at the same
location. The trajectories are remarkably similar, despite the wide range
of initial cloud density contrasts.}
\end{figure}

\begin{figure}
\plotone{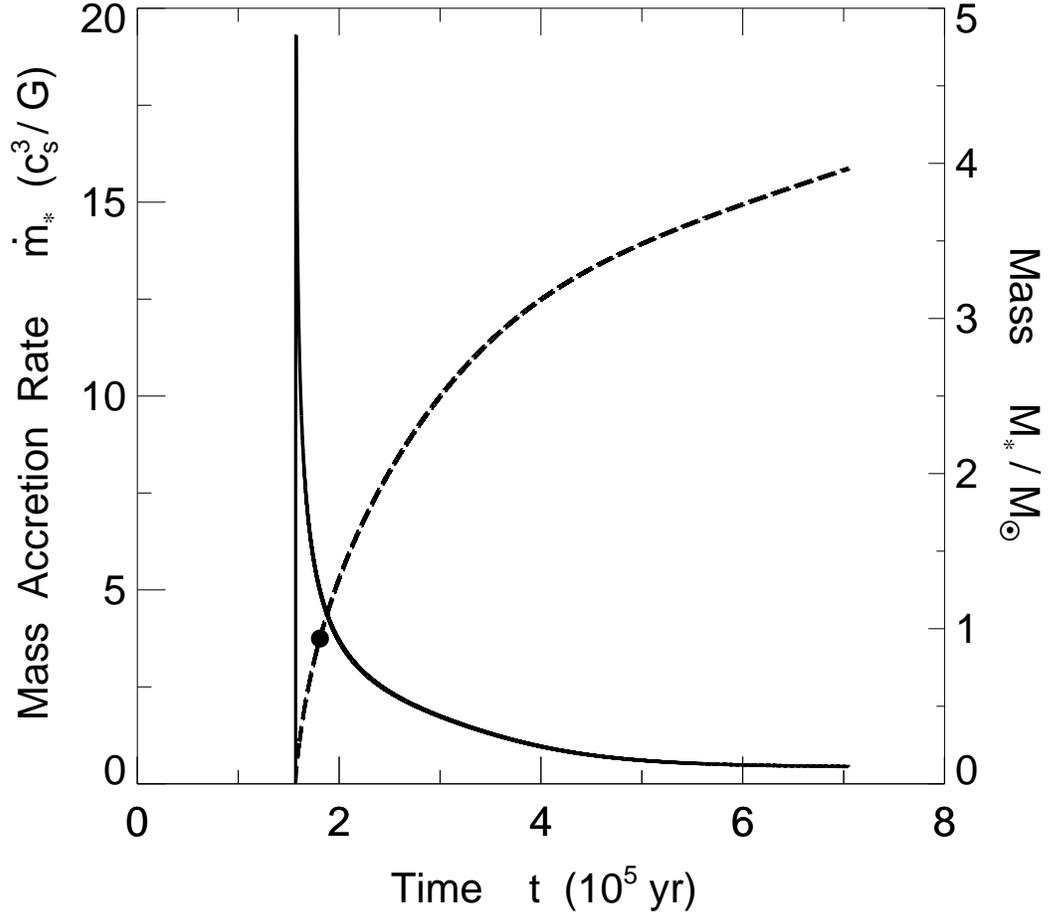}
\caption{Evolution of the mass accretion rate ({\it solid curve}) and stellar
mass ({\it dashed curve}) for the cloud with an initial density contrast
of \hbox{$\rho_c/\rho_b\,=\,100$}. The filled circle marks the time $t_f$.} 
\end{figure}

\end{document}